\documentclass[twocolumn,showpacs,aps]{revtex4}
\usepackage{amssymb}

\usepackage{graphicx}
\usepackage{dcolumn}
\usepackage{bm}

\begin{document}
\title {Thermally activated energy and critical magnetic fields of  SmFeAsO$_{0.9}$F$_{0.1}$}
\author {Y. Z. Zhang}
\author{Z. A. Ren}
\author{Z. X. Zhao}

\affiliation{National Laboratory for superconductivity, Institute of
Physics and Center for Condensed Matter Physics, Chinese Academy of
Sciences, P. O. Box 603,  100190, Beijing, China}


\begin{abstract}

Thermally activated flux flow and vortex glass transition of
recently discovered SmFeAsO$_{0.9}$F$_{0.1}$ superconductor are
studied in magnetic fields up to 9.0 T. The thermally activated
energy is analyzed in two analytic methods, of which one is
conventional and generally used, while the other is closer to the
theoretical description. The thermally activated energy values
determined from both methods are discussed and compared. In
addition, several critical magnetic fields determined from
resistivity measurements are presented and discussed.

\end{abstract}

 \maketitle

\section{Introduction}

The recently discovered FeAs-based superconductors inspired study,
as their superconducting transition temperatures and upper critical
magnetic fields reach the values which are higher than those of
MgB$_2$ and comparable to those of the cuprate based superconductors
(CBS) \cite{Kamihara, Ren, Ren2, Wen, Chen, Chen2, Cruz, Dong,
Hunte, Senatore, Jia}. One of interesting characteristics is that
they show layered structure with conducting layers of FeAs and
charge reservoir layers of ReO, where Re is a rare earth element
\cite{Ren2}. This layered structure is very similar to that of CBS,
and suggests that the superconducting behaviors may have
similarities to those of CBS.

The vortex dynamics of CBS have been widely studied in theories and
experiments \cite{Yeshurun, Malozemoff, Hagen, Geshkenbein, Kes,
Vinokur, Blatter, Brandt, Shi, Tinkham, Palstra, Zhang3, Zhang2,
Figueras, Xiaowen, Zhang, Andersson, Ravelosona, Gordeev, Yang}.
According to the theory, the thermally activated flux flow (TAFF)
resistivity is expressed as $\rho =
(2\nu_0LB/J)\exp(-J_{c0}BVL/T)\sinh(JBVL/T)$ \cite{Blatter, Brandt,
Shi, Tinkham, Palstra}, where $\nu_0$ is an attempt frequency for  a
flux bundle hopping, $L$ the hopping distance, $B$ the magnetic
induction, $J$ the applied current density, $J_{c0}$ the critical
current density in the absence of  flux creep,  $V$ the bundle
volume, and $T$ the temperature. If $J$ is small enough and
$JBVL/T\ll 1$, we have
\begin{equation}
\rho= (2\rho_cU/T)\exp(-U/T), \label{eq1}
\end{equation}
where $U=J_{c0}BVL$ is the thermally activated energy (TAE),  and
$\rho_c=\nu_0LB/J_{c0}$. Equation (\ref{eq1}) simply means that the
prefactor $2\rho_cU/T$ is temperature and magnetic field dependent.

Generally, the TAE of CBS is analyzed by equation (\ref{eq1}) using
an assumption that the prefactor $2\rho_cU/T$ is temperature
independent, and $\ln \rho(H,T)$ linearly depends on $1/T$ with the
form $\ln \rho(H,T)=\ln\rho_0(H)-U_0(H)/T$, where $H $ is the
magnetic field strength, $\ln \rho_0(H)=\ln\rho_{0f}+U_0(H)/T_c$
[note that $\ln \rho_0(H)$ is the $\ln \rho(H)$ value for $1/T\to
0$], $\rho_{0f}$ the constant, $U_0$ the TAE for $T\to 0$, and $T_c$
the superconducting transition temperature. The importance is that
the analysis leads to $U=U_0(1-t)$, and the apparent activate energy
$-\partial \ln\rho/\partial T^{-1} =U_0$, where $t=T/T_c$
\cite{Zhang3,Zhang2, Palstra, Figueras, Xiaowen, Zhang, Andersson,
Ravelosona, Gordeev, Yang}. By drawing resistivity data in the
so-called Arrhenius plot with a relation $\ln \rho(H,1/T)$, one can
easily determine $U_0(H)$ with its corresponding slope in a low
resistivity range. However , $U=U_0(1-t)$ may not be true in
reality, when a local slope of $\ln \rho$ vs $1/T$ in the Arrhenius
plot shows a round curvature. As a result, the corresponding
apparent activated energy $-\partial \ln \rho(H,T)/\partial T^{-1}$
shows a sharp increase with decreasing temperature. The phenomena
means $U\ne U_0(1-t)$, $\rho_{0f}\ne$ const, and the determination
of $U_0(H)$ from the slopes is a problem. In the early stage of the
discovery of CBS, the experimental observation of the abnormal
phenomena had been reported by Palstra {\it et al} \cite{Palstra}
without solution.

Zhang {\it et al} \cite{Zhang3} suggested that the temperature
dependent of the prefactor in equation (\ref{eq1}) must be taken
into account in the analysis. By using this suggestion to equation
(\ref{eq1}) with $U=U_0(1-t)^q$, the derivative,
\begin{eqnarray}
-\partial \ln \rho/\partial T^{-1}= (1-T/U)(U-T\partial U/\partial
T)\cr    =[U_0(1-t)^q-T] [1+q t/(1-t)], \label{eq2}
\end{eqnarray}
where $q$ has a value in the range from 0.5 to 2, and $t=T/T_c$.

In this paper, TAFF resistivity of SmFeAsO$_{0.9}$F$_{0.1}$ (SFAOF)
is studied with magnetic fields up to 9.0 T. By using the assumption
$\rho_{0f}=2\rho_cU/T=$ const, and $U=U_0(1-t)$,  the TAFF behaviors
were first analyzed in the Arrhenius plot. After that,  by using the
assumption that the prefactor $2\rho_cU/T$ is temperature dependent,
and $\rho_c$ is not temperature dependent, the TAFF resistivity is
analyzed in the other method in which equation (\ref{eq2}) is
employed to determine $U_0(H)$. The $U_0(H)$ determinations from
both methods are discussed and compared.  We suggest that the second
method shall be instead of the first in the analysis of TAFF
characteristics of other superconductors. In addition, the vortex
glass transition, zero resistivity temperature, and the temperature
dependents of the critical fields determined from different
resistivity criteria values in the superconducting transition regime
are presented.

\section{Experiments}

The SFAOF sample was prepared by a high-pressure  synthesis method.
SmAs powder (pre-sintered) and As, Fe, Fe$_2$O$_3$, FeF$_2$ powders
(the purities of all starting chemicals are better than 99.99\%)
were mixed together with the nominal stoichiometric ratio of
Sm[O$_{1-x}$F$_x$]FeAs, then ground thoroughly and pressed into
small pellets. The pellet was sealed in boron nitride crucibles and
sintered in a high pressure synthesis apparatus under the pressure
of 6.0 GPa and temperature of 1250$^\circ$ C for two hours. The
x-ray diffraction analysis showed that the main phase is LaOFeAs
structure with some impurity phases \cite{Ren, Ren2}. It was cut
into a rectangular shape with dimensions of 4.20 mm(length)$\times$
1.60 mm(width) $\times$ 1.08 mm(thickness). The standard four probe
technique was used  for resistivity $\rho(T,H)$ measurements.
Bipolar pulsed dc current with an amplitude of 5.0 mA (corresponding
to the current density about $\sim$0.29 A/cm$^2$) was applied to it.
The measurements were performed on a physical property measurement
system (PPMS, Quantum Design) with the magnetic field up to 9 T.
From  the zero field $\rho(T,0)$ data, we find that the
superconducting transition width is about 1.7 K (defined by the
superconducting transition from 10\% to 90\% of the normal
resistivity), and the zero resistant temperature $T_{c0}$ is 52.2 K
(determined by the criterion of  0.1 ${\mu\Omega \cdot}$cm).

\section{Results and discussion}


\begin{figure}
\includegraphics
[width= .85\columnwidth] {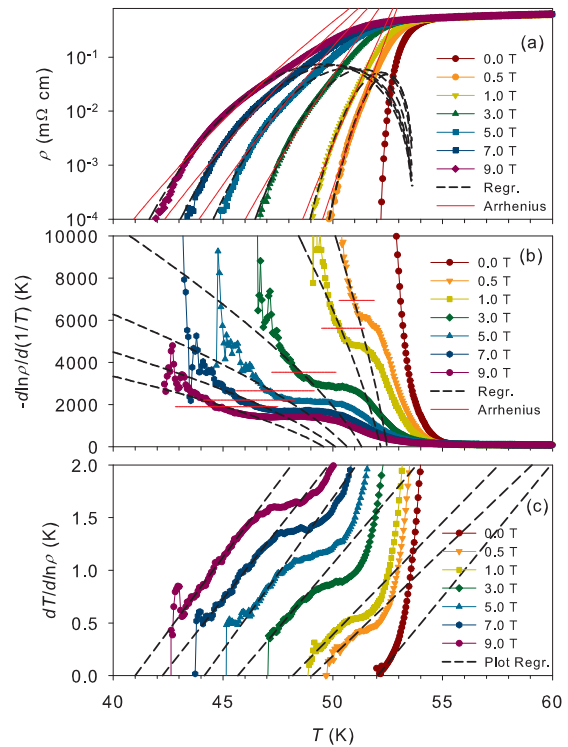} \caption{(a)-(c), respectively,
show $\rho(T,H)$, $-\partial \ln \rho(T,H)/\partial T^{-1}$, and
$\partial T/\partial \ln \rho(T,H)$ data in magnetic fields of
$\mu_0H=0.0$, 0.5,  1.0, 3.0, 5.0, 7.0, and  9.0 T with different
symbols. The solid lines in (a) and short horizontal lines in (b)
correspond to the regressions with the fitting parameters of
$U_0(H)$ and $\rho_{0f}$ determined from the first analytic method.
The dashed lines are regression curves with the fitting parameters
$U_0(H)$ and $\rho_c(H)$ determinded from the second analytic method
(see text). The dashed lines in (c) plot linear fittings.
\label{f1}}
\end{figure}


Figure \ref{f1}(a)-(c), respectively, show  $\rho(T,H)$, $-\partial
\ln \rho(T,H)/\partial T^{-1}$, and $\partial T/\ln\rho(T,H)$ data
with different symbols for 0.0, 0.5, 1.0, 3.0, 5.0, 7.0, and 9.0 T
fields.  One shall notice that the suppression of the
superconductivity at 9.0 T field is comparable to that of low
anisotropic CBS, e.g. YBa$_2$Cu$_3$O$_{7-\delta}$ (YBCO), and
suggests that the anisotropy is similar to YBCO. The so-called
apparent activated energy $-\partial \ln \rho(T,H)/\partial T^{-1}$
data were generally used to analyze TAE \cite{Palstra,Zhang3}, and
the derivative $\partial T/\ln\rho(T,H)$ data is always used to
analysis the vortex glass transition boundary \cite{Fisher,
Fisher2,Wagner, Zhang1}. Below, we will present our analysis.

\begin{figure}
\includegraphics
[width= .85\columnwidth] {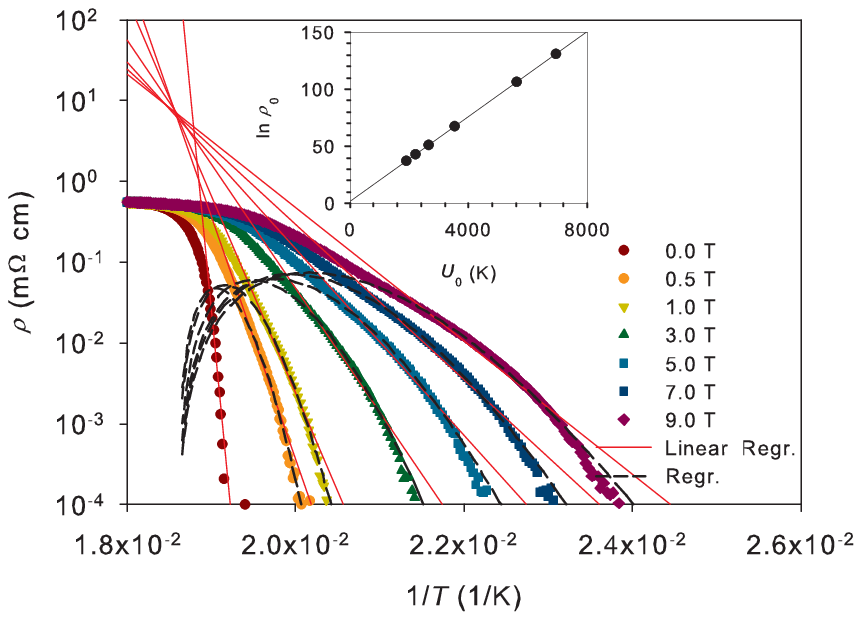} \caption{Arrhenius plot of
$\rho(T,H)$ for $\mu_0H=0.0$, 0.5,  1.0, 3.0, 5.0, 7.0, and
 9.0 T. The solid lines are plot linear fitting for low
resistivity. The dashed lines are regressive curves using the second
analytic method. The inset shows $\ln \rho_0(U_0)$ data determined
by using the first analytic method, where the solid line is a linear
fit for the data.\label{f2}}
\end{figure}

Figure \ref{f2} shows the Arrhenius plot of its resistivity in
magnetic fields of $\mu_0H=0.0$, 0.5,  1.0, 3.0, 5.0, 7.0, and  9.0
T with different symbols. The solid lines plot linear regressions in
the low resistivity range. According to the conventional analysis
method, these linear fits are based on the assumption
$U(T,H)=U_0(H)(1-t)$, where each $U_0(H)$ was determined by using
each slope. Note that there is a cross point where all the linear
fitting lines approximately focus together except for the line for
zero field, and this point leads to $T_{cr} \approx 53.8$ K. The
inset shows $\ln \rho_0(U_0)$ data and the corresponding solid line
with the linear fitting $\ln \rho_0=\ln\rho_{0f}+U_0/T_c$. From the
fitting, we determined $\rho_{0f}=6.668$ m$\Omega\cdot$cm, and
$T_c\approx 53.8$ K which coincides with the value of $T_{cr}$.
Using the $\rho_{0f}$, $T_c$, and $U_0(H)$ data, we regressed the
$\rho(T,H)$ data as shown in figure \ref{f1}(a)  with solid lines.
Since the assumption of $U(T,H)=U_0(H)(1-t)$ leads to $-\partial \ln
\rho(T,H)/\partial T^{-1}=U_0$, we present $U_0(H)$ data in figure
\ref{f1}(b) with horizontal solid lines where each of them has a
limited length. Each length covers the temperature interval which
corresponds to the interval of the reciprocal temperature for
determining $U_0(H)$ in  the Arrhenius plot. For decreasing
temperature, note that each $-\partial \ln \rho(T,H)/\partial
T^{-1}$ dataset approximately intersects its horizontal $U_0(H)$
line center with a divergent trend in the temperature interval. This
means that each $U_0(H)$ approximates to the average value of its
$-\partial \ln \rho(T,H)/\partial T^{-1}$ in the temperature
interval. The similar divergent trends were also observed in YBCO
\cite{Palstra, Zhang3}. The analysis indicates that the TAE
determined from the conventional method may have problems in
characteristics, and the method suggested by Zhang {\it et al}
\cite{Zhang3} ought to be taken into consideration.

In figure \ref{f1}(b), one may have noticed that each $-\partial \ln
\rho(T,H)/\partial T^{-1}$ set can be divided into five regimes: (i)
the normal state regime at high temperature, where $-\partial \ln
\rho(T,H)/\partial T^{-1}$ is almost temperature-and
magnetic-field-independent; (ii) the primary superconducting
transition regime, where  $-\partial \ln \rho(T,H)/\partial T^{-1}$
quickly increases and the resistivity begins sharply decreasing (see
figure \ref{f1}(a)); (iii) the platform regime, where $-\partial \ln
\rho(T,H)/\partial T^{-1}$ for each field measurement show a step
structure (see figure \ref{f1}(b)); (iv) the second sharply
increasing regime, where $-\partial \ln \rho(T,H)/\partial T^{-1}$
quickly increases into a high value range and $\partial T/\partial
\ln \rho$ data show a linear type (see figure \ref{f1}(c)); (v) the
strong fluctuation regime, where $-\partial \ln \rho(T,H)/\partial
T^{-1}$ curve shows an irregular shape due to the resistivity
reaching the low measurable range.

For analyzing TAFF behaviors with the second analytic method, the
first thing is to validate one or two regimes which relate to the
TAFF behaviors. Apparently, the data in regimes (i) and (ii) do not
relate to TAFF behaviors, as the data are in the normal state in
regime (i) and in the flux flow regime in regime (ii). The data show
a platform structure and possibly relate to TAFF behaviors in regime
(iii). Note that resistivity data in the regime are only about one
order of magnitude less than that in regime (i) (where the
resistivity is in the normal state) [see figure \ref{f1}(a) and
(b)]. Therefore, we conclude that the data in regime (iii) are not
in the TAFF regime \cite{Palstra}. In regime (iv), the resistivity
is about 2 and 3 orders of magnitude less than that in regime (i).
The resistivity in this range, as suggested by Palstra {\it et al}
\cite{Palstra}, relates to TAFF behavior.  According to the vortex
glass transition theory, the linear (ohmic) resistivity ought to
linearly vanish in a form $\rho \propto(T-T_g)$, where $T_g$ is the
vortex glass transition temperature \cite{Fisher, Fisher2}.
Accordingly, $\partial T/\partial\ln\rho\propto (T-T_g)$
\cite{Wagner, Zhang1}. In figure \ref{f1}(b), one will easily find
that $\partial T/\partial\ln\rho(T,H)$ curves show approximately
linear curvatures in regime (iv). The data in the regime (v) show
irregular trend, as the resistivity is going to a deeply
superconducting state where the resistivity may be dominated by
non-ohmic characteristics; besides, the measuring voltmeter reached
its low measuring limitation in experiments. The analysis concludes
that the TAFF resistivity is in regime (iv).

\begin{figure}
\includegraphics
[width= 0.85\columnwidth] {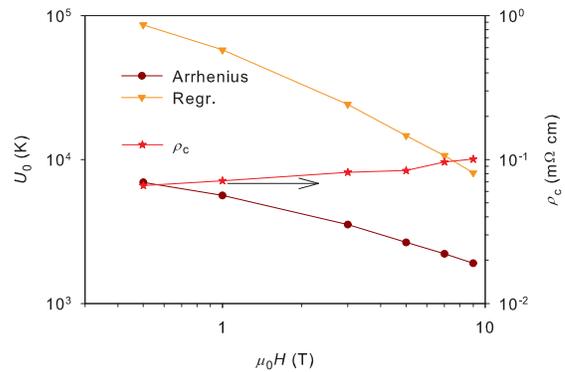} \caption{(a) The circles show
$U_0(H)$ extracted from the slopes in Arrhenius plot in figure
\ref{f2}. The downward triangles show $U_0(H)$ data determined by
using the relation $-\partial \ln \rho/\partial
T^{-1}=[U_0(1-t)^q-T][1+qt/(1-t)]$, where $q=2$. The stars represent
corresponding $\rho_c$ data. \label{f3}}
\end{figure}

Figure \ref{f3} shows $U_0(H)$ determined by the first analytic
method with circles and by the second with triangles. The stars show
corresponding $\rho_c(H)$ for the second method. All the dash lines
in figures \ref{f1}(a) and (b), and figure \ref{f2} are regression
curves with the regression parameters of $U_0(H)$ and $\rho_c(H)$ in
figure \ref{f3}. In the analysis of these data with the second
method, we first derived $U_0$ with equation (\ref{eq2}) as $U_0$ is
the only free parameter in the equation except for $q$. We found
that the energy relation, $U(T,H)=U_0(1-t)^q$ with $q=2$, leads to
good consistency to the experimental data, where $t=T/T_c$, and
$T_c=53.8$ K. After determination of each $U_0(H)$, each $\rho_c(H)$
can be easily determined by fitting equation (\ref{eq1}). One will
find that all the regressions (dashed lines) in figure \ref{f1}(a),
figure \ref{f1}(b), and figure \ref{f2} are in good agreement with
experimental data and confirm the correctness of the analytic
method. Note that, although each temperature interval in regime
(iv), which we used to determine $U_0(H)$ with the second analytic
method, is somewhat less than that we used to determine $U_0(H)$
with the first analytic method (see figure \ref{f1}(b) and (c)), the
regressions of the second method are still given better fitting
results.

For magnetic field above 1.0 T, we find that $U\propto
H^{-0.57}(1-t)$ for the data derived from Arrhenius plot with the
first method, and $U\propto H^{-0.99}(1-t)^2$ with the second
method. Note that the TAE determined by the second method is about
one order larger than that determined from the first as shown in the
figure. The reason of the large difference between the two analytic
methods is due to that the first method not taking into account the
temperature dependent relation of the prefactor in equation
(\ref{eq1}), while the second method takes into consideration.
Apparently, the second method is  closer to the theoretical
description.

\begin{figure}
\includegraphics
[width= 0.85\columnwidth] {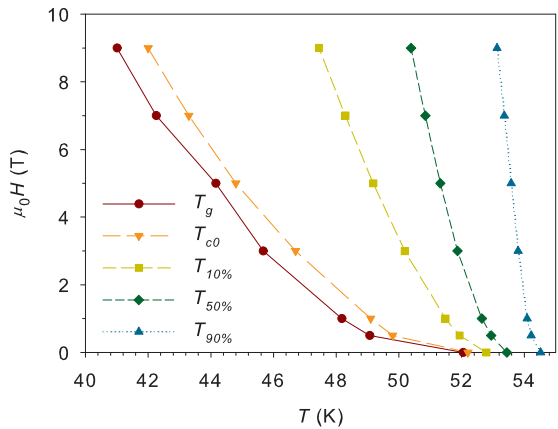} \caption{Critical magnetic
fields. The circles show the $H(T_g)$ data, the downward triangles
the $H(T_{c0})$, the squares the $H(T_{10\%})$, the diamonds the
$H(T_{50\%})$, and the upward triangles the $H(T_{90\%})$.
\label{f4}}
\end{figure}

Figure \ref{f4} shows the vortex glass transition boundary $H(T_g)$,
$H(T_{c0})$, $H(T_{10\%})$, $H(T_{50\%})$, and $H(T_{90\%})$,where
$T_g$ is the glass transition temperature, $T_{c0}$ the zero
resistance temperature, $T_{10\%}$ the 10\% of the normal state
resistivity temperature, $T_{50\%}$ the 50\% and $T_{90\%}$ the
90\%. Here, $H(T_g)$ was determined by using the $\partial
T/\partial\ln\rho\propto (T-T_g)$ relation [dash lines in figure
\ref{f2}(c)]. We find that the data can be fitted by a relation
$H_g=H_{g0}(1-T_g/52.05)^2$, where $\mu_0H_{g0}\approx 201$ T. The
$\mu_0H_{g0}$ value may not be true in zero temperature, but it
leads to a conclusion that the vortex melting field is very large
for SFAOF superconductor when the temperature approaches to zero. A
similar characteristic can be also observed for $H(T_{c0})$ data,
since each $T_{c0}(H)$ is slightly higher than that $T_g(H)$. The
normalized transition widths (defined as $(1-T_{10\%}/T_{90\%})$ are
comparable to YBCO, and thus suggest that the anisotropy of the
SFAOF is similar to YBCO. The $H(T_{90\%})$ curve shows a steep
increase indicating that the upper critical magnetic field of the
superconductor is very high in low temperature. One may notice that
the transition widths between $T_{c0}(H)$ and $T_{10\%}(H)$ are
comparable to that between $T_{10\%}(H)$ and $T_{90\%}(H)$,
suggesting that superconductor may have better analysis results when
further improving the quality of the superconductor.

In comparisons, a sample that was prepared in the same sample batch
was used in ac susceptibility measurements with magnetic fields up
to 7.0 T, and frequencies up to 1.11 kHz. We found that $T_g(H)$ is
somewhat larger than the corresponding temperature of the peak in
the imaginary component in ac susceptibility measurement. However,
the similar $H$ increasing trends were also found in the ac
measurement. Besides, we found that the onset temperature of the
superconducting transition $T_{onset}(H)$ of ac susceptibility
measurement is somewhat higher than that of $T_{c0}(H)$. Detailed
analysis of the ac measurement is just ongoing.

\section{Conclusion}

In summary, we analyze the resistive TAFF behavior of SFAOF with two
theoretical analysis. For the first method, $\rho_{0f}=$ const and
$U=U_0(1-t)$ were assumed, and thus Arrhenius plot was employed in
analysis. This method is simple and easy in analysis, but the
analysis results remain inconsistency to experimental data. The
second method assumes that the prefactor $2\rho_cU/T$ is temperature
dependent, while $\rho_c$ is not temperature dependent. By using the
second method,  equation (\ref{eq2}) is obtained.  The second method
results in the regressions are in good agreement with experimental
data. The TAE analysis shows $U\sim H^{-1}(1-t)^2$ for magnetic
field above 1.0 T. The second method is also simple and easy, since
only two free parameters in analysis.  We suggest that the second
method shall be instead of the first for the analysis of TAFF
behaviors of other superconductors. The study shows that critical
magnetic fields of SFAOF may have large values in low temperature
and comparable to CBS.

\section{Acknowledgements}

We are grateful to Prof. H. H. Wen and Dr. H. Yang for resistivity
measurements and helpful discussion. This work has been financially
supported by the National Natural Science Foundation of China (Grant
No. 10874221).


\begin{thebibliography}{}
\bibitem{Kamihara} Y. Kamihara, T. Watanabe, M. Hirano, and H. Hosono, J. Am. Chem.
Soc. 130, 3296 (2008).
\bibitem{Ren} Z. A. Ren, W. Lu, J. Yang, W. Yi, X. L. Shen, Z. C. Li, G. C.
Che, X. L. Dong, L. L. Sun, F. Zhou, and Z. X. Zhao, Chin. Phys.
Lett. 25, 2215 (2008).
\bibitem{Ren2} Z. A. Ren, G. C. Che, X. L. Dong, J. Yang, W. Lu, W. Yi, X. L. Shen,
Z. C. Li, L. L. Sun, F. Zhou, Z. X. Zhao, Europhysics Letters  83,
17002 (2008).
\bibitem{Wen} H. H. Wen, G. Mu, L. Fang, H. Yang, and X. Zhu, Europhys. Lett. 82, 17009 (2008).
\bibitem{Chen}  G. F. Chen, Z. Li, D. Wu, G. Li, W. Z. Hu, J. Dong, P. Zheng, J. L. Luo, N. L. Wang , Phys. Rev. Lett. 100, 247002 (2008).
\bibitem{Chen2}  G. F. Chen, Z. Li, D. Wu, J. Dong, G. Li, W. Z. Hu, P. Zheng, J. L. Luo, N. L. Wang, Chin. Phys. Lett. 25, 2235
(2008).
\bibitem{Cruz} Clarina de la Cruz, Q. Huang, J. W. Lynn,
Jiying Li, W. Ratcliff II, J. L. Zarestky, H. A. Mook, G. F. Chen,
J. L. Luo, N. L. Wang, Pengcheng Dai, Nature 453, 899 (2008).
\bibitem{Dong} J. Dong, H. J. Zhang, G. Xu, Z. Li, G. Li, W. Z. Hu, D. Wu,
G. F. Chen, X. Dai, J. L. Luo, Z. Fang, N. L. Wang, Europhysics
Letters, 83, 27006 (2008).
\bibitem{Hunte} F. Hunte, J. Jaroszynski, A. Gurevich, D. C. Larbalestier, R. Jin,
A. S. Sefat, M. A. McGuire, B. C. Sales, D. K. Christen, and D.
Mandrus, Nature (London) 453, 903 (2008).
\bibitem{Senatore} C. Senatore, M. Cantoni, G. Wu, R.H. Liu, X.H.
Chen, and R. Fl\"ukiger, arXiv:0805.2389.
\bibitem{Jia} Y. Jia, P.
Cheng, L. Fang, H. Q. Luo, H. Yang, C. Ren, L. Shan, C. Z. Gu, and
H. H. Wen, arXiv:0806.0532.
\bibitem{Yeshurun} Y. Yeshurun and A. P. Malozemoff, Phys. Rev. Lett. {\bf 60}, 2202 (1988).
\bibitem{Malozemoff} A. P. Malozemoff, L. Krusin-Elbaum, D. C. Cronemeyer, Y. Yeshurun and F. Holtzberg, Phys. Rev. B {\bf 38}, 6490 (1988).
\bibitem{Hagen} C. W. Hagen and R. Griessen, Phys. Rev. Lett. {\bf 62}, 2857 (1989).
\bibitem{Geshkenbein} V. B. Geshkenbein, M. V. Feigel\rq man, A. I. Larkin, and V. M. Vinokur, Physica C \textbf{162}, 239 (1989).
\bibitem{Kes} P. H. Kes, J. Aarts, .J van den Berg, C. J. van der Beek and J. A. Mydosh, Supercond. Sci. Technol. \textbf {1}, 242 (1989).
\bibitem{Vinokur}V. M. Vinokur, M. V. Feigel'man, V. B. Geshkenbein, and A. I. Larkin, Phys. Rev. Lett. \textbf{65}, 259
(1990);  J. Kierfeld, H. Nordborg, and V. M. Vinokur, Phys. Rev.
Lett. \textbf{85}, 4948 (2000).
\bibitem{Blatter} G. Blatter, M. V. Feigel\rq man, V. B. Geshkenbein, A. I. Larkin, and V. M. Vinokur, Rev. Mod.
Phys. \textbf{66}, 1125 (1994), (and the references therein).
\bibitem{Brandt}E. H. Brandt,  Rep. Prog. Phys. {\bf 58}, 1465 (1995), (and the references therein).
\bibitem{Shi} D. Shi, High-Temperature Superconducting Materials Science and Engineering New Concepts and Technology,  (Pergamon, New York, 1995).
\bibitem{Tinkham} M. Tinkham, Introduct. to Superconduct., McGraw-Hill, New York, 1996, (and the references therein).
\bibitem{Palstra} T. T. M. Palstra, B. Batlogg, L. F. Schneemeyer and J. V. Waszczak, Phys. Rev. Lett. 61, 1662 (1988);
T. T. M. Palstra, B. Batlogg, R. B. van Dover, I. F. Schneemeyer,
and J. V. Waszczak, Phys. Rev. B \textbf{41}, 6621 (1990).
\bibitem{Zhang3} Y. Z. Zhang, H. H. Wen, Z. Wang, Phys. Rev. B \textbf{74},  144521
(2006).
\bibitem{Zhang2}Y. Z. Zhang, Z. Wang, X. F. Lu, H. H. Wen, J. F. de
Marneffe, R. Deltour, A. G. M. Jansen, and P. Wyder, Phys. Rev. B
\textbf{71},  052502 (2005).
\bibitem{Figueras} J. Figueras, T. Puig, and X. Obradors
Phys. Rev. B {\bf 67}, 014503 (2003)
\bibitem{Xiaowen} Cao Xiaowen, Wang Zhihe, and Xu Xiaojun
Phys. Rev. B {\bf 65}, 064521 (2002); Xu Xiaojun, Fu Lan, Wang
Liangbin, Zhang Yuheng, Fang Jun, Cao Xiaowen, Li Kebin, and Sekine
Hisashi Phys. Rev. B {\bf 59}, 608 (1999).
\bibitem{Zhang} Y. Z. Zhang, R. Deltour, and Z. X. Zhao, Phys. Rev. Lett. {\bf
85}, 3492 (2000).
\bibitem{Andersson} M. Andersson, A. Rydh, and \"O. Rapp
Phys. Rev. B {\bf 63}, 184511 (2001).
\bibitem{Ravelosona} D. Ravelosona, J. P. Contour, and N. Bontemps
Phys. Rev. B {\bf 61}, 7044 (2000).
\bibitem{Gordeev} S. N. Gordeev, A. P. Rassau, R. M. Langan, P. A. J. de Groot, V. B. Geshkenbein, R. Gagnon, and L. Taillefer Phys. Rev. B {\bf 60}, 10477 (1999).
\bibitem{Yang} H. C. Yang, L. M. Wang, and H. E. Horng
Phys. Rev. B {\bf 59}, 8956 (1999).
\bibitem{Fisher} M.P.A. Fisher, Phys. Rev. Lett. 62, 1415 (1989).
\bibitem{Fisher2} D.S. Fisher, M.P.A. Fisher, and D.A. Huse, Phys. Rev. B
43, 130 (1991).
\bibitem{Wagner} P. Wagner, U. Frey, F. Hillmer, and H. Adrian,
Phys. Rev. B 51, 1206 (1995).
\bibitem{Zhang1} Y. Z. Zhang, R. Deltour, J.-F. de Marneffe, Y. L. Qin, L. Li,
Z. X. Zhao, A. G. M. Jansen, and P. Wyder Phys. Rev. B {\bf 61},
8675 (2000).

\end{thebibliography}
\end{document}